\documentclass[intlimits,twoside,a4paper]{article}

\usepackage[cp1251]{inputenc}

\usepackage[eqsecnum]{cmpj3}

\usepackage{bm}

\issue{2020}{23}{2}{23605}
\doinumber{10.5488/CMP.23.23605}

\title[Diffusion of hard sphere fluids in disordered porous media: Enskog theory description
]{Diffusion of hard sphere fluids in disordered porous media: Enskog theory description\footnote{Dedicated to Ihor Mryglod on the ocassion of his 60th birthday.}

\author{M.F. Holovko,  M.Ya. Korvatska }
\address{
Institute for Condensed Matter Physics of the National
Academy of Sciences of Ukraine,\\ 1 Svientsitskii St., 79011 Lviv,
Ukraine
}}
\date{Received February  27, 2020, in final form March 13, 2020}

\begin{document}

\maketitle

\begin{abstract}
	We use the Enskog theory for the description of the self-diffusion coefficient of  hard sphere fluids in disordered porous media. Using the scaled particle theory previously developed by us for the description of thermodynamic properties of hard sphere fluids, simple analytical expressions for the contact values of the fluid-fluid and fluid-matrix pair distribution functions are obtained and used as the input of Enskog theory. The expressions obtained for the contact values are described only by the geometric porosity and do not include the dependence on other types of porosity that are important for the description of thermodynamic properties. It is shown that the application of such contact values neglects the effects of trapping of fluid particles by a matrix and at least the probe particle porosity $\phi$ should be included in the Enskog theory for a correct description of the matrix influence. In this paper we extend the Enskog theory by changing the contact values of the fluid-matrix and the fluid-fluid pair distribution functions with new properties which include the dependence  not only on geometric porosity but also on probe particle porosity $\phi$. It is shown that such semi-empirical improvement of the Enskog theory corresponds to SPT2b1 approximation for the description of thermodynamic properties and it predicts correct trends for the influence of porous media on the diffusion coefficient of a hard sphere fluid in disordered porous media. Good agreement with computer simulations is illustrated. The effects of fluid density, fluid to matrix sphere size ratio, matrix porosity and matrix morphology on the self-diffusion coefficient of hard sphere fluids are discussed.
	\keywords  
	hard sphere fluid, disordered porous media,  Enskog theory, self-diffusion coefficient, scaled particle theory, probe particle porosity

%	\pacs 61.43.Gt,66.10.Cb
\end{abstract}

\section{Introduction}

The properties of fluids confined in disordered porous media substantially differ  from those of bulk fluids \cite{GelbGub99}. For the last three decades, starting from the pioneering work of Madden and Glandt \cite{MadGlandt88}, great theoretical efforts have been devoted to the study of fluid adsorption in disordered porous media. So far most of these theoretical and simulation-based studies have been focused on the static structural and thermodynamic properties and only a few of them are devoted to investigations of dynamic properties~\cite{ChangJad04,Krakov,KurCos11}. According to the model proposed by Madden and Glandt \cite{MadGlandt88}, a  disordered porous medium is presented as a  matrix of quenched configurations of randomly distributed particles. The specificity of such an approach is connected with the double quenched-annealed averages: the annealed average is taken over all fluid configurations while the additional quenched average should be taken over all realizations of the matrix. A standard approach to solve this problem is based on the replica method. Using the replica Ornstein-Zernike (ROZ) integral equation theory \cite{GivenStall92}, the statistical mechanics approach of liquid state was extended to a  description of different models of fluids confined in random porous media \cite{Ros99,Pizio00} including the chemical reacting fluids adsorbed in porous media \cite {Trokh96,Trokh97}. However, unlike bulk fluids, no analytical results have been obtained in this approach even for the simplest model such as a hard sphere fluid in a hard-sphere matrix.

%Dedicated to Ihor Mryglod on the ocassion of his 60th birthday.

In order to solve this problem, Holovko and Dong \cite{HolDong09} proposed to extend the classical scaled particle theory (SPT) \cite{ReissFrisLeb59,ReissFrisHel60} for the description of thermodynamic properties of hard sphere fluids in disordered porous media. During the last decade the SPT approach for hard sphere fluids in disordered porous media was essentially improved and developed \cite{ChenDong10,HolShmot10,PatHol11,HolPat12,HolPatDong12,HolPatDong17}. The approach proposed in  \cite{HolDong09} and named SPT1, contains a subtle inconsistency appearing when the size of matrix particles is considerably larger than the size of fluid particles. This inconsistency was eliminated in a new approach named SPT2 \cite{PatHol11}. Consequently, the first rather accurate analytical expressions were obtained for the chemical potential and pressure of a hard sphere fluid confined in a hard sphere (HS) or of an overlapping hard sphere (OHS) matrix. The obtained expressions include three parameters defining the porosity of the matrix. The first one is related to the bare geometry of the matrix. It is the so-called geometric porosity $\phi_{0}$ characterizing the free volume, which is not occupied by matrix particles. The second parameter $\phi$ is defined by the chemical potential of a fluid in the limit of infinite dilution. It is  the so-called probe particle porosity characterizing the adsorption of a fluid in an empty matrix. Usually, $\phi\leqslant\phi_{0}$. The third parameter $\phi^{*}$ is defined by the maximum value of the fluid packing fraction of a hard sphere fluid in a porous medium. It characterizes the maximum adsorption capacity of a matrix for a given fluid. It is responsible for crowding effects of a fluid in porous media. In \cite{HolPatDong12}, a general expression for $\phi^{*}$ was proposed 
\begin{equation}
\frac{1}{\phi^{*}}=\left(\frac{1}{\phi}-\frac{1}{\phi_{0}}\right)\left[\ln\left(\frac{\phi_{0}}{\phi}\right)\right]^{-1},
\end{equation}
which is exact for one-dimensional case and can be considered a good approximation for higher dimensions.

The developed SPT2 approach was generalized for fluids of anisotropic particles \cite{HolShmot14,HolShmot18},  for a hard sphere mixture \cite{ChenZhao16} and for a mixture of hard sphere and anisotropic particles \cite{HvozdPat18,HolPat15} in disordered porous media. The obtained results for the models considered were also used  as the reference system taking into account attractive \cite{HolPat15,HolShmot20}, associative \cite{KalHol14} and Coulombic \cite{HolPatPat16,HolPatPat17,HolovPatPat17} interactions.

The SPT2 approach is also useful  for the description of static structure of hard sphere fluids in disordered porous media. In particular, in \cite{KalHol14} a simple and rather accurate expression was obtained for the contact value of the fluid-fluid pair distribution function  which was used for the description of phase behaviour and percolation properties of the patchy colloidal fluids in disordered porous media. The structural information provided by the SPT2 approach similar to the bulk case can be used as a source of input in the Enskog theory for the description of transport properties of hard sphere fluids in porous media (for the bulk case see the reference \cite{ResibLee77} and references therein for details).

The first attempt to apply the Enskog theory for a hard sphere fluid in a hard-sphere matrix was considered by Yethiraj with coworkers \cite{ChangJad04} for the self-diffusion coefficient. They considered a fluid in a disordered matrix as a mixture of two components one of which is quenched in space and was treated as the particles with the infinite mass. In such an approach, the contact values of fluid-fluid and fluid-matrix pair-distribution functions are introduced as the input of theory. In \cite{ChangJad04}, they were taken from  computer simulations. The comparison between theoretical and simulation results shows a systematic overestimation of theoretical predictions which increase with the increasing fraction of matrix particles. It is clear that in order to improve the theoretical predictions we should modify the contact values of pair distribution functions. In \cite{KalHol14}, the contact value of the fluid-fluid pair distribution function was obtained in a simple analytical form and was found to be in good agreement with computer simulations. A similar expression can also be found for the contact value of the fluid-matrix pair distribution function. However, in contrast to thermodynamic properties,  the obtained expressions for the contact values of both pair distribution functions are described only by the geometric porosity $\phi_{0}$ and do not include the dependence either on the probe particle porosity $\phi$ or on the porosity $\phi^{*}$. We  show in this paper that the  application of such contact values neglects the effects of fluid particles being trapped by the matrix, and at least the probe particle porosity $\phi$ should be included in the Enskog theory for a correct description of the matrix effect.

In this report, we propose such an improvement in a  manner similar to the description of thermodynamic properties. We  show that such semi-empirical improvement of the Enskog equation predicts correct trends for the effect of porous media on the diffusion coefficient of hard sphere fluids in disordered porous media. The theory developed is applied to the investigation of the diffusion of hard sphere fluids in different porous media, namely to a hard sphere matrix and to an overlapping hard sphere matrix. The effects of fluid density, matrix porosity, matrix morphology, and fluid to matrix sphere size ratio on the self-diffusion coefficient will be discussed. We also present some comparison with computer simulation data obtained by the group of Yethiraj \cite{ChangJad04}.

\section{The Enskog theory}

The Enskog theory is based on the assumption that each collision between hard spheres is completely independent and instantaneous \cite{ResibLee77}. In theory, only binary collision is considered and all higher multiple collisions are neglected. Moreover, it is assumed that the frequency of the binary collision increases by an amount  proportional to the probability of one particle finding its neighbours. The motion of particles of type $\mu$ in the fluid can be described in terms of the velocity autocorrelation function
\begin{equation}
\psi_{\mu}(t)=\left\langle \vartheta_{\mu}(t)\vartheta_{\mu}(0)\right\rangle,
\label{HolKor2.1}
\end{equation}
where $\vartheta_{\mu}(t)$ is the velocity of the particle type $\mu$ at time $t$ and $\langle \ldots\rangle $ is the equilibrium ensemble average. The time evolution of $\psi_{\mu}(t)$ is well described by the generalized Langevin equation \cite{Boon80}, where the role of memory kernel is played by the friction coefficient $\xi_{\mu}(t)$. The frequency dependent $\xi_{\mu}(t)$ is given by the Green-Kubo formula \cite{Boon80}
\begin{equation}
\xi_{\mu}(z)=\frac{1}{3kT}\int_{0}^{\infty}\rd t {\re}^{-zt}\left\langle  F_{\mu}(t)F_{\mu}(0)\right\rangle ,
\label{HolKor2.2}
\end{equation}
where $F_{\mu}(t)$ is the force between a fixed particle of type $\mu$ and the surrounding particles at a time  $t$, $k$ is the Boltzmann constant, $T$ is the absolute temperature. We also note that  $\langle F_{\mu}(t)F_{\mu}(0)\rangle $ is not the usual force-force time correlation function since the time evolution of $F(t)$ is given by a Liouville projection operator
\begin{equation}
F_{\mu}(t)=\exp\left( \ri QL_Nt\right) F_{\mu}(0),
\label{HolKor2.3}
\end{equation}
where $\ri L_N$ is the Liouville operator for an $N$-particle system,  $Q=1-\mathcal{P}$ and $\mathcal{P}$ is the Mori-Zwanzig projection operator defined as \cite{Boon80}
\begin{equation}
\mathcal{P}*=\left\langle *\vartheta_{x}\right\rangle \vartheta_{x}/\langle \vartheta_{x}^{2}\rangle .
\label{HolKor2.4}
\end{equation}
The self-diffusion coefficient $D_{\mu}$ for particles of type $\mu$ is related to the corresponding friction coefficient $\xi_{\mu}$ via the Einstein relation as
\begin{equation}
D_{\mu}= k T/\xi_{\mu}\,,  
\label{HolKor2.5}
\end{equation}
where $\xi_{\mu}=\xi_{\mu}(z=0)$ is the friction coefficient in the stationary limit.

For a mixture of hard spheres, the Enskog theory leads to the following expression \cite{ChangJad04,McQuarrie76}
\begin{equation}
\xi_{\mu}=32 \left(\frac{k Tm_{\mu}}{2\piup\sigma_{\mu\mu}^{2}}\right)^{1/2}\sum_{\nu}\left( 1+\frac{m_{\mu}}{m_{\nu}}\right)^{-1/2}\eta_{\nu}
\frac{\sigma^{2}_{\mu\nu}}{\sigma^{2}_{\nu\nu}}
\frac{\sigma_{\mu\mu}}{\sigma_{\nu\nu}}g_{\mu\nu}(\sigma_{\mu\nu}),  
\label{HolKor2.6}
\end{equation}
where $m_{\mu}$, $m_{\nu}$ are the masses of particles of types $\mu$ and $\nu$, respectively, $\eta_{\nu}={1}/{6}\piup\rho_{\nu}\sigma_{\nu\nu}^{3}$ is the packing fraction of particles of type $\nu$, $\rho_{\nu}$ is the number density,
\begin{equation}
\sigma_{\mu\nu}=\frac12\left( \sigma_{\mu\mu}+\sigma_{\nu\nu}\right) 
\label{HolKor2.7}
\end{equation}
is the hard sphere interaction diameter between species $\mu$ and $\nu$, and $g_{\mu\nu}(\sigma_{\mu\nu})$ is the pair distribution function at contact between species $\mu$ and $\nu$.

Similar to \cite{ChangJad04}, we  mimic a hard sphere fluid in a hard sphere matrix by a binary mixture where one component is infinitely massive. As a result, we have
\begin{equation}
\xi_{1}=32\left(\frac{k T m_{1}}{2\piup\sigma_{11}^{2}}\right)^{1/2}\left[ \frac{1}{\sqrt{2}}\eta_{1}g_{11}(\sigma_{11})+\frac14\tau(\tau+1)^{2}\eta_{0}g_{10}(\sigma_{10})\right]. 
\label{HolKor2.8}
\end{equation}
Hereafter,  we use the conventional notations \cite{GivenStall92,Ros99,Pizio00}, where the index ``1'' is used to denote the fluid component and the index ``0'' denotes the matrix particle, $\tau=\sigma_{11}/\sigma_{00}$.

In accordance with (\ref{HolKor2.5}), we have the following expression for self-diffusion of a hard sphere fluid in a disordered porous medium 
\begin{equation}
{D}_{1}/D_{1}^{0}=\frac{\sqrt{2\piup}}{32} \left[ \frac{1}{\sqrt{2}}\eta_{1}g_{11}(\sigma_{11})+\frac14\tau(\tau+1)^{2}\eta_{0}g_{10}(\sigma_{10})\right]^{-1} ,
\label{HolKor2.9}
\end{equation}
where $D_{1}^{0}= ({k T\sigma_{11}^{2}}/{m_{1}}) ^{1/2}$.

For further calculations, the expressions for the contact values of the pair distribution functions $g_{11}(\sigma_{11})$ and $g_{10}(\sigma_{10})$ are needed. To this end, we  develop and improve the SPT2 approach presented in \cite{KalHol14}. We start from $g_{11}(\sigma_{11})$. In accordance with the SPT2 approach, the contact value of a small scaled particle and a fluid particle can be presented in the form
\begin{equation}
g_{1\text{s}}(\sigma_{1\text{s}})=1\big/\left[ p_{0\text{s}}(\lambda_\text{s})-\eta_{1}(1+\lambda_\text{s})^{3}\right], 
\label{HolKor2.10}
\end{equation}
where $\sigma_{1\text{s}}=1/2(\sigma_{11}+\sigma_\text{ss})$, $\lambda_\text{s}={\sigma_\text{ss}}/{\sigma_{11}}$, $p_{0\text{s}}(\lambda_\text{s})$ is the probability of finding a cavity created by the scaled  particle in the matrix in the absence of fluid \cite{HolPat12}. Here, the index ``s'' is used for the scaled particle. For the hard sphere matrix
\begin{equation}
p_{0\text{s}}(\lambda_\text{s})=1-\eta_{0}(1+\tau\lambda_\text{s})^{3}  .
\label{HolKor2.11}
\end{equation}
%
%where $k_{10}=\sigma_{11}/\sigma_{00}$.

In order to obtain the expression for $g_{11}(\sigma_{11})$, we expand $g_{1\text{s}}(\sigma_{1\text{s}})$ as
\begin{equation}
g_{1\text{s}}(\sigma_{1\text{s}})=G_{1\text{s}}^{(0)}+G_{1\text{s}}^{(1)}\frac{\lambda_\text{s}}{1+\lambda_\text{s}}+\frac12G_{1\text{s}}^{(2)} \frac{\lambda_\text{s}^{2}}{(1+\lambda_\text{s})^{2}} \,,
\label{HolKor2.12}
\end{equation}
where $G_{1\text{s}}^{(0)}$, $G_{1\text{s}}^{(1)}$ and $G_{1\text{s}}^{(2)}$ are found from the continuity of $g_{1\text{s}}(\sigma_{1\text{s}})$ and the first and second derivatives with respect to $\lambda_\text{s}$ at $\lambda_\text{s}=0$. Subsequently, we can put in (\ref{HolKor2.12}) $\lambda_\text{s}=1$. As a result, we  have
\begin{equation}
g_{11}(\sigma_{11})=\frac{1}{\phi_{0}-\eta_{1}}+\frac32\frac{\eta_{1}+\eta_{0}\tau}{(\phi_{0}-\eta_{1})^{2}}+\frac12\frac{(\eta_{1}+\eta_{0}\tau)^{2}}{(\phi_{0}-\eta_{1})^{3}} \,, 
\label{HolKor2.13}
\end{equation}
where $\phi_{0}=1-\eta_{0}$ is the geometric porosity. We note that in the calculation $G_{1\text{s}}^{(2)}$ the derivative was taken only from the dominator of the first derivative from (\ref{HolKor2.10}) and the numerical coefficient $\frac94$ was changed to $\frac12$ in order to describe correctly Carnahan-Starling correction \cite{HolShmot18}.

The expression for the contact value $g_{10}(\sigma_{10})$ can be found in a similar manner and can be presented in the form 
\begin{equation}
g_{10}(\sigma_{10})=\frac{1}{\phi_{0}-\eta_{1}}+\frac{3}{1+\tau}\frac{\eta_{1}+\eta_{0}\tau}{(\phi_{0}-\eta_{1})^{2}}+\frac{2}{(1+\tau)^{2}}\frac{(\eta_{1}+\tau\eta_{0})^{2}}{(\phi_{0}-\eta_{1})^{3}} . 
\label{HolKor2.14}
\end{equation}
%%%%%%%%%%%%%%%%%%%%%%%%%%%%%%%%%%%%%%%%%%%%%%%%%%
\begin{figure}[!t]
	%\begin{center}
		\begin{minipage}[h]{0.45\linewidth}
			\includegraphics[width=1\textwidth]{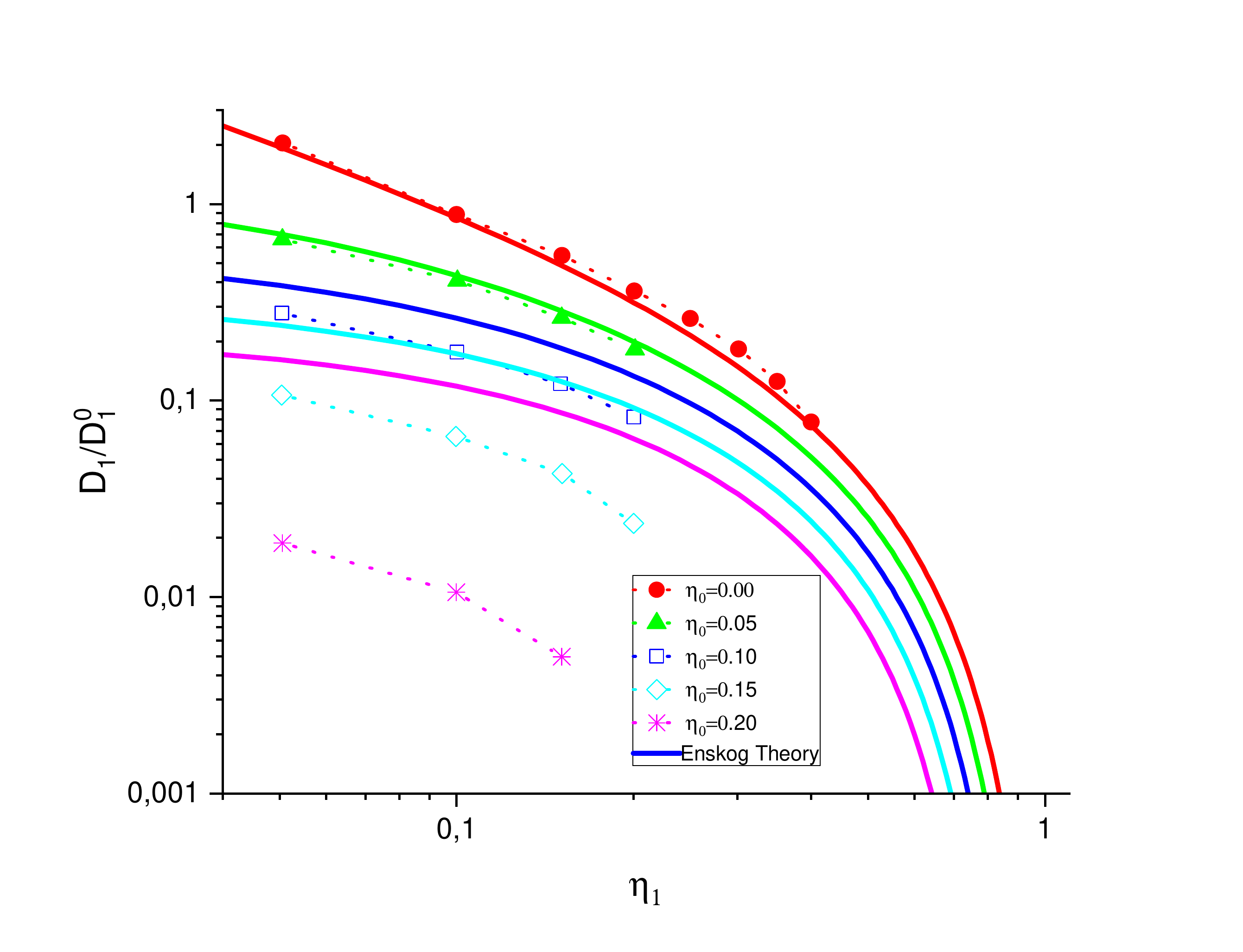}
			\caption{(Colour online) Comparison of the standard Enskog theory prediction and computer simulation results for the self-diffusion coefficient $D_{1}$ of a hard-sphere fluid in a hard-sphere matrix as a function of the fluid packing fraction $\eta_{1}$ for different matrix packing fractions $\eta_{0}$ and for $\tau=1$.} %% подпись к рисунку
			\label{FIGHoLKor1} 
		\end{minipage}
		\hfill
				\begin{minipage}[h]{0.45\linewidth}
						\vspace{-12ex}
		\includegraphics[width=1\textwidth]{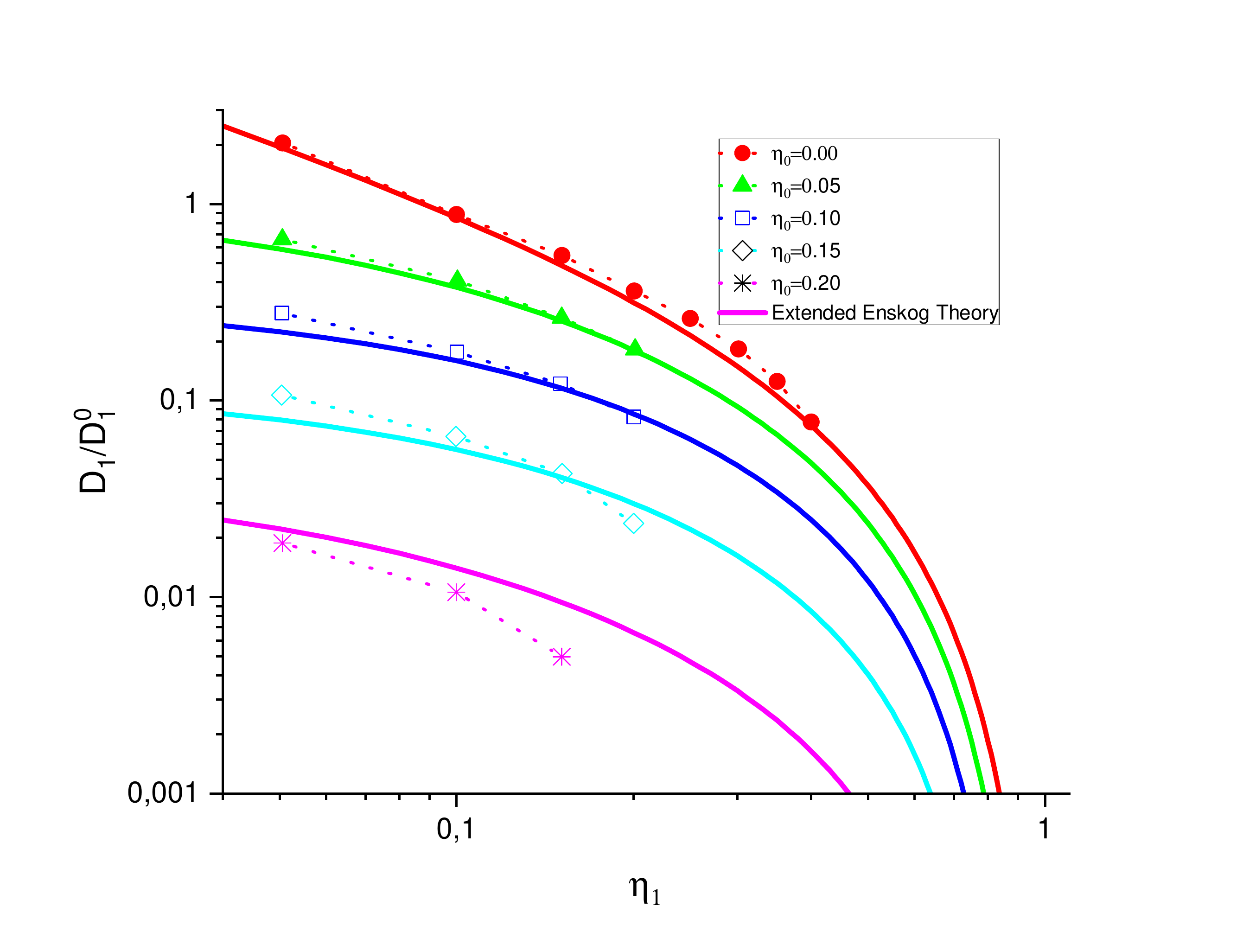}
			\caption{(Colour online) Same as in figure~\ref{FIGHoLKor1} for the extended Enskog theory.}
			\label{FIGHoLKor2}
		\end{minipage}
	%\end{center}
\end{figure}
%%%%%%%%%%%%%%%%%%%%%%%%%%%%%%%%%%%%%%%%%%%%%%%%%
%
Figure~\ref{FIGHoLKor1} depicts the prediction from the Enskog theory versus computer simulation data taken from \cite{ChangJad04} for the dependence of the self-diffusion coefficient $D_{1}$ on the fluid packing fraction $\eta_{1}$ for the case $\tau=1$ at different values of the packing fraction of matrix particles $\eta_{0}$. As we can see, with an increasing $\eta_{1}$ or $\eta_{0}$, the self-diffusion coefficient $D_{1}$ monotonously decreases. We have a very good agreement between the theory and computer simulations for the bulk case and for a rather low value $\eta_{0}=0.05$. However, similar to~\cite{ChangJad04} figure~\ref{FIGHoLKor1} demonstrates that with an increasing $\eta_{0}$, the theory greately overestimates the value of $D_{1}$ by an order of magnitude for $\eta_{0}=0.20$.
Yethiraj  with coworkers \cite{ChangJad04} noted that such a discrepancy between the Enskog theory prediction and computer simulation results demonstrates that with an increasing $\eta_{0}$, configurations of the fluid and the disordered matrix are very different from those of the equilibrium mixtures, which is the basic assumption in the Enskog theory. We remark that in the equilibrium case, the presence of matrix particles hinders the dynamics of fluids only by the geometric porosity $\phi_{0}=1-\eta_{0}$, which according to (\ref{HolKor2.13}) and (\ref{HolKor2.14}) defines the contact values $g_{11}(\sigma_{11})$ and $g_{10}(\sigma_{10})$. However, in the case of immobile particles of the matrix, there are additional effects from the matrix due to the geometric constrains induced by the obstacles, which is not present in the equilibrium case. These additional effects become  significant with an increasing $\eta_{0}$, slowing down the translation diffusion of fluids. In this paper we  show that this effect is strongly connected with the probe particle porosity $\phi$ introduced by us for the description of thermodynamic properties of hard sphere fluids \cite{PatHol11,HolPat12}, which, however, is not present in the description of contact values of the pair distribution functions $g_{11}(\sigma_{11})$ and $g_{10}(\sigma_{10})$.

\section{Revision and extension of the Enskog theory to hard sphere fluids in disordered porous media}

We remember that the original Enskog equation for a hard sphere fluid was formulated nearly a hundred years ago as the generalization of the Boltzmann kinetic equation to high densities of hard sphere fluids \cite{Boon80,Chapman70}. It includes, as a multiplier the pressure term $\left( p_{1}/k T\rho_{1}-1\right) $, which due to the virial theorem \cite{YukhHol80} 
\begin{equation}
\frac{ p_{1}}{k T\rho_{1}}-1=4\eta_{1}g_{11}(\sigma_{11})  
\label{HolKor3.1}
\end{equation}
 can be presented in a well-known form \cite{ResibLee77,McQuarrie76} via the contact value of the fluid-fluid distribution function $g_{11}(\sigma_{11})$, while the expression for self-diffusion of a hard sphere fluid can be presented in the well-known form
\begin{equation}
{D_{1}}/{D_{1}^{0}}=\frac{\sqrt{\piup}}{16}\left[ \eta_{1}g_{11}(\sigma_{11})\right]^{-1}  .
\label{HolKor3.2}
\end{equation}
However, for a hard sphere fluid in a porous medium, no simple virial expressions like (\ref{HolKor3.1}) for the pressure $p_{1}$ exist and it is somewhat problematic to write an expression like (\ref{HolKor2.9}) for self-diffusion of a hard sphere fluid in a disordered porous medium. Of course, we can write the expression~(\ref{HolKor2.9}) for the $D_{1}/D_{1}^{0}$  but in this expression we cannot consider $g_{11}(\sigma_{11})$ and $g_{10}(\sigma_{10})$ as the contact values of the fluid-fluid and fluid-matrix distribution functions. In general, they are some thermodynamic properties defined by equation~(\ref{HolKor2.9}). In this paper, we do not modify the expression (\ref{HolKor2.13}) for $g_{11}(\sigma_{11})$. We  consider and modify only the expression (\ref{HolKor2.14}) for $g_{10}(\sigma_{10})$. We start from the infinite dilution. When $\eta_{1}\to0$
\begin{equation}
{D_{1}}/{D_{1}^{0}}\to \frac{\sqrt{2\piup}}{32}\left[ \frac14\tau(\tau+1)^{2}\eta_{0}g_{10}(\sigma_{10},\eta_{1}=0)\right]^{-1} .  
\label{HolKor3.3}
\end{equation}
In the equilibrium case, according to (\ref{HolKor2.14})
\begin{equation}
g_{10}(\sigma_{10},\eta_{1}=0)=\frac{1}{\phi_{0}}\left[ 1+\frac{3}{1+\tau}\frac{\eta_{0}}{\phi_{0}}\tau+\frac{2}{(1+\tau)^{2}}\left( \frac{\eta_{0}}{\phi_{0}}\right) ^{2}\tau^{2}\right]. 
\label{HolKor3.4}
\end{equation}

In this paper, we change the expression (\ref{HolKor3.4}) to
\begin{equation}
g_{10}(\sigma_{10},\eta_{1}=0)=\frac{\phi_{0}}{\phi} \,,
\label{HolKor3.5}
\end{equation}
which can be considered as the ratio of the probability of finding a cavity created by a fluid  particle in an empty matrix to the probability of finding a cavity created by a point particle in an empty matrix. $\phi_{0}$ and $\phi$ are the geometric porosity and the probe particle porosity, respectively. In accordance with \cite{HolPat12}, for a hard sphere fluid in a hard sphere matrix
\begin{equation}
\phi_{0}=1-\eta_{0}\,,\quad
\phi=\phi_{0}\exp\left\lbrace  -\frac{\eta_{0}\tau}{1-\eta_{0}}\left[ 3(1+\tau)+\frac92\tau\frac{\eta_{0}}{1-\eta_{0}}+\frac{1+\eta_{0}+\eta_{0}^{2}}{(1-\eta_{0})^{2}}\tau^{2}\right] \right\rbrace.   
\label{HolKor3.6}
\end{equation}

In the presence of fluid particles, similar to thermodynamic properties \cite{HolPat12,HolPatDong12,HolPatDong17}, we  put
\begin{equation}
g_{10}(\sigma_{10})=\frac{\phi_{0}}{\phi}\left[\frac{1}{1-\eta_{1}/\phi}+\frac{3}{1+\tau}\frac{{\eta_1}/{\phi_{0}}}{\left( 1-\eta_{1}/\phi_{0}\right) ^{2}}
+\frac{2}{(1+\tau)^{2}}\frac{({\eta_{1}}/{\phi_{0}})^{2}}{\left( 1-\eta_{1}/\phi_{0}\right) ^{3}}\right] .
\label{HolKor3.7}
\end{equation}

The first term in (\ref{HolKor3.7}) leads to a divergence at $\eta_{1}=\phi$ and similar to thermodynamic consideration, we change $1/(1-\eta_{1}/\phi)$ to \cite{HolPat12,HolPatDong12,HolPatDong17}
\begin{equation}
\frac{1}{1-\eta_{1}/\phi}\to
 \frac{1}{1-\eta_{1}/\phi_{0}}+ \frac{\eta_{1}(\phi_{0}-\phi)}{\phi_{0}\phi(1-\eta_{1}/\phi_{0})^{2}}.
\label{HolKor3.8}
\end{equation}
Consequently, for $g_{10}(\sigma_{10})$, we  have
\begin{equation}
g_{10}(\sigma_{10})=\frac{\phi_{0}}{\phi}\left[ 
\frac{1}{1-\eta_{1}/\phi_{0}}+ \frac{\eta_{1}(\phi_{0}-\phi)}{\phi_{0}\phi(1-\eta_{1}/\phi_{0})^{2}}
+\frac{3}{1+\tau}\frac{\eta_{1}/\phi_{0}}{(1-\eta_{1}/\phi_{0})^{2}}+ \frac{2}{(1+\tau)^{2}}\frac{(\eta_{1}/\phi_{0})^{2}}{(1-\eta_{1}/\phi_{0})^{3}}\right] .
\label{HolKor3.9}
\end{equation}
Now, for the diffusion coefficient we again have the expression (\ref{HolKor2.9}) in which,  however, $g_{10}(\sigma_{10})$ is given by (\ref{HolKor3.9})
and for $g_{11}(\sigma_{11})$ we have the previous form (\ref{HolKor2.13}). The results obtained from this expression for $\tau=1$ are presented in figure~\ref{FIGHoLKor2}.
As we can see, the improved version of the Enskog theory is much better than the standard version (see figure~\ref{FIGHoLKor1}). It agrees with the computer simulation data and quite accurately reproduces the trend of change in the dependence of matrix particles on the packing fraction.

\begin{figure}[!b]
	\centerline{
		\includegraphics [height=0.42\textwidth]{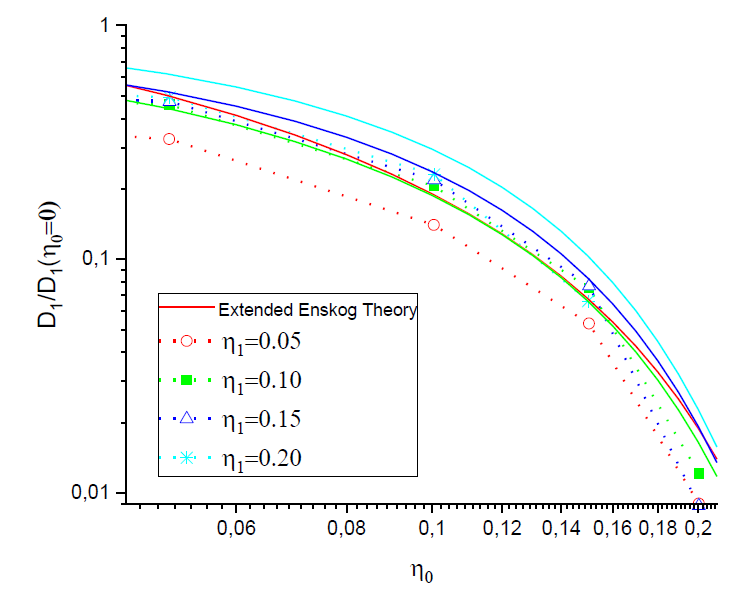}
	}
	\caption{ (Colour online) Comparison of the extended Enskog theory prediction and computer simulation results \cite{ChangJad04} for the diffusion coefficient $D_{1}$ for a hard sphere fluid in a hard sphere matrix normalized by the diffusion coefficient $D_{1}(\eta_{0}=0)$ of the hard sphere fluid of the same size without matrix as a function of the matrix packing fraction $\eta_{0}$ for a different fluid packing fraction $\eta_{1}$ and for $\tau=1$.}
	\label{FIGHoLKor3}
\end{figure}
In figure~\ref{FIGHoLKor3}, we compare the prediction for the ratio $D_{1}/D_{1}(\eta_{0}=0)$ as a function of $\eta_{0}$ against computer simulations results from \cite{ChangJad04}, where $D_{1}(\eta_{0}=0)$ is the diffusion of the fluid at the same value of $\eta_{1}$ as for $D_{1}$ but with $\eta_{0}=0$. As we can see, after extension, the Enskog theory  reproduces the trends of the computer simulation results. The dependence of $D_{1}$ on $\eta_{0}$ is more or less similar for all values $\eta_{1}$. Only our theory does not reproduce the effects of underestimation of $D_{1}/D_{1}(\eta_{0}=0)$ for $\eta_{1}=0.05$ observed in computer simulations \cite{ChangJad04} and slightly overestimates the values $D_{1}/D_{1}(\eta_{0}=0)$ at a higher fluid density $\eta_{1}=0.20.$
%%%%%%%
%
\begin{figure}[!t]
	\centerline{
		\includegraphics [height=0.42\textwidth]{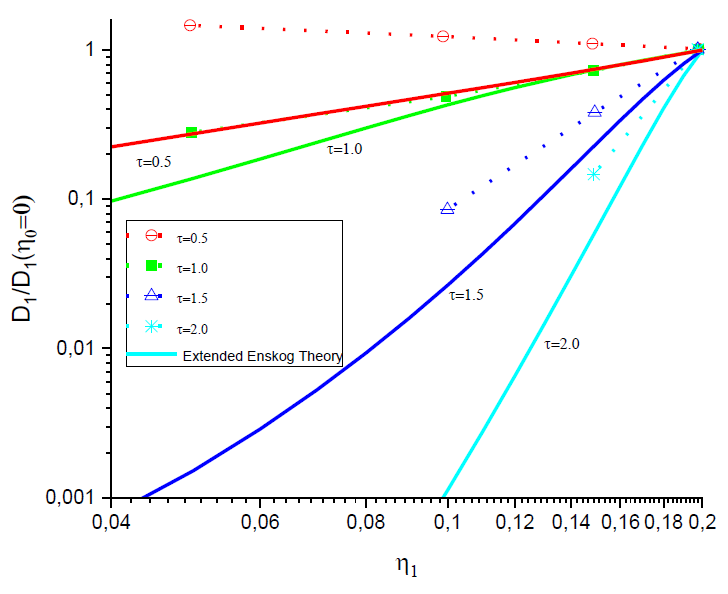}
	}
	\caption{(Colour online) The extended Enskog theory prediction versus computer simulation results \cite{ChangJad04} for the diffusion coefficient $D_{1}$ for a hard sphere fluid in a hard sphere matrix normalized by the diffusion coefficient $D_{1}(\eta_{0}=0)$ of the hard sphere fluid of the same size without matrix as a function of the fluid packing fraction $\eta_{1}$ at a fixed value $\tau$. The total packing fraction $\eta=\eta_{0}+\eta_{1}=0.20$ for all cases. }
	\label{FIGHoLKor4}
\end{figure}

Now, we discuss the effect of the fluid to matrix sphere size ratio $\tau$ on the self-diffusion coefficient of a hard sphere fluid in a disordered porous medium. Figure~\ref{FIGHoLKor4} illustrates the diffusion coefficient of a hard sphere fluid in a disordered porous medium normalized by the diffusion coefficient of hard spheres of the same size in the absence of the matrix, $D_{1}/D_{1}(\eta_{0}=0)$, as a  function of $\eta_{1}$ for different $\tau$. Similar to \cite{ChangJad04}, the total packing fraction $\eta=\eta_{1}+\eta_{0}$ is fixed at $\eta=0.20$ for all cases. For comparison, in figure~\ref{FIGHoLKor4} the computer simulation data taken from \cite{ChangJad04} are presented as well. As we can see, theoretical and simulations results are different. In all cases, the theoretical predictions are lower than the computer simulations data. However, both approaches qualitatively show the same trend of dependence on the size ratio which is opposite to the theoretical results obtained from the standard Enskog theory \cite{ChangJad04}. 
In the limiting case, when $\tau\to0$, $D_{1}\to D_{1}( {\eta_{1}}/{\phi_{0}})$ and the ratio $D_{1}/D_{1}(\eta_{0}=0)$ cannot be larger than unity contrary to the simulation data \cite{ChangJad04}.  In the opposite case, when $\tau\to\infty$ and $\eta_{0}\to0$ in such a way that $\tau^{3}\eta_{0}=1/6\piup\rho_{0}\sigma_{11}^{3}$ is fixed \cite{HolPatDong17}, we  have
\begin{equation}
\frac{D_{1}}{D_{1}(\eta_{0}=0)}=\left\lbrace  1+\frac14\eta_{0}\tau^{3}\frac{\sqrt{2}}{\eta_{1}}\frac{1}{\phi}\left[ 1+\frac{1-\phi}{\phi}\frac{D_{1}(\eta_{0}=0)}{D_{1}^{0}}\right] \right\rbrace  ^{-1}<1,
\label{HolKor3.10}
\end{equation}
where
\begin{equation}
\phi=\exp\left(  -\eta_{0}\tau^{3}\right)  .
\label{HolKor3.11}
\end{equation}

As we can see, the ratio $D_{1}/D_{1}(\eta_{0}=0)$ decreases with a decreasing $\eta_{1}$ and with an increasing $\tau$. As noted by Yethiraj with coworkers \cite{ChangJad04}, the size-dependent behaviour of diffusion of a fluid in a porous medium can be qualitatively  explained using the concept of free volume \cite{Minton}. Replacing a big fluid particle by several immobile matrix particles of smaller sizes increases the  excluded volume for the particles and this slows down the translation motion of the fluid particles. In the opposite case, replacing small fluid hard spheres by an immobile particle of a bigger size decreases the excluded volume for the diffusing particles. As a result, diffusion of fluid particles becomes faster.

Finally we discuss the effect of morphology of a porous medium on the behaviour of self-diffusion coefficient of hard sphere fluids. To this end, we consider a hard sphere fluid in two different matrices, namely a hard sphere matrix and an overlapping hard sphere matrix. For the self-diffusion coefficient of a hard sphere fluid in a porous medium, we use the expression (\ref{HolKor2.9}), where $g_{11}(\sigma_{11})$ and $g_{10}(\sigma_{10})$ are given by the expression (\ref{HolKor2.13}) and (\ref{HolKor3.9}), respectively. The porosities $\phi$ and $\phi_{0}$ for a hard sphere matrix are given in (\ref{HolKor3.6}) and for the overlapping hard sphere matrix in \cite{HolPat12} 

\begin{equation}
 \phi=\exp\left[ -(1+\tau)^{3}\eta_{0}\right]  \quad\text{and}\quad \phi_{0}=\exp(-\eta_{0}).
\label{HolKor3.12}
\end{equation}

\begin{figure}[!t]
	\centerline{
		\includegraphics [height=0.42\textwidth]{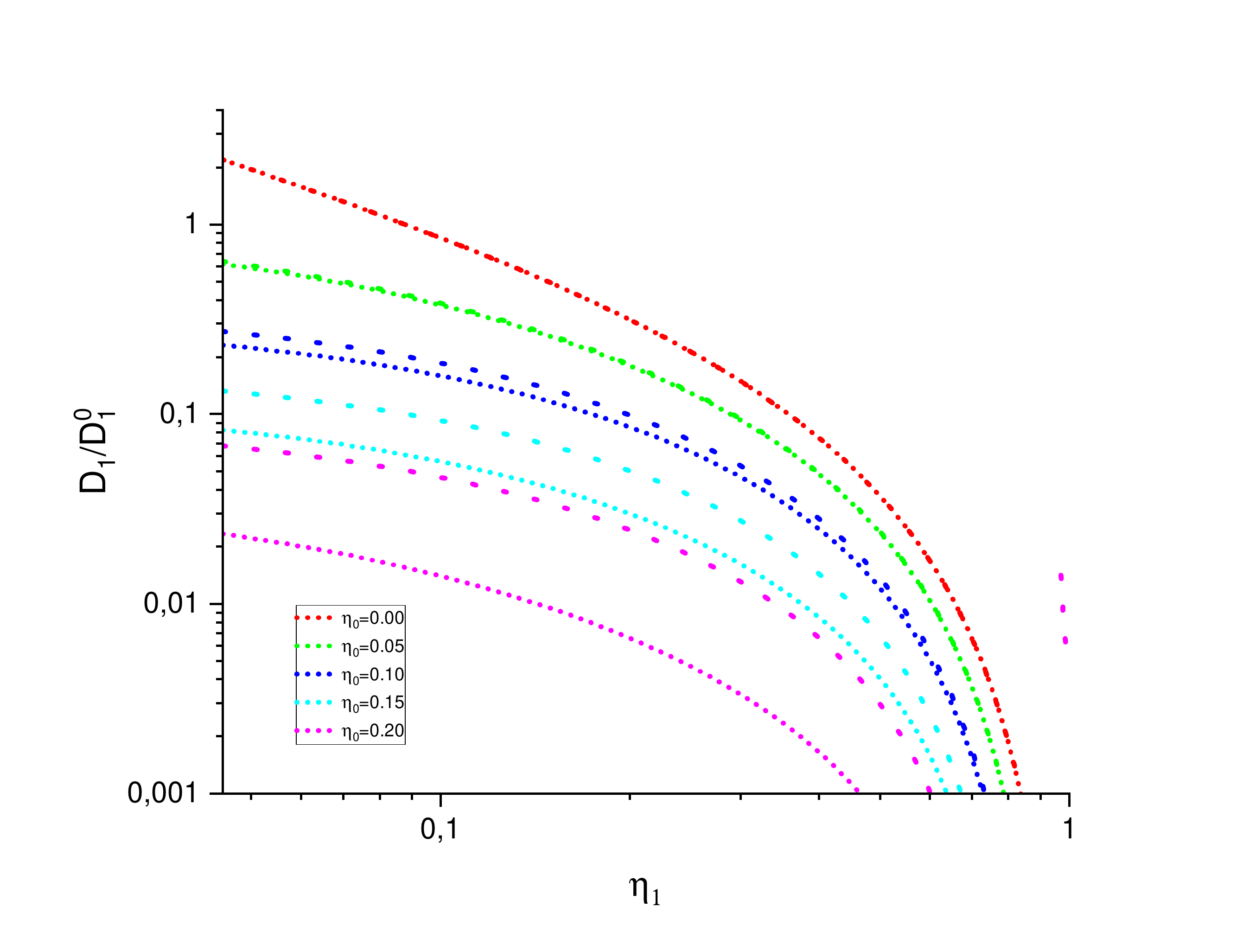}
	}
	\caption{(Colour online) The self-diffusion coefficient $D_{1}$ of a hard sphere fluid calculated from the extended Enskog theory in a hard sphere matrix and in an overlapping hard sphere matrix as a function of the fluid packing fraction $\eta_{1}$ for different matrix packing fractions $\eta_{0}$ and for $\tau=1$. The dotted line corresponds to the hard sphere matrix and the dashed line corresponds to the overlapping hard sphere matrix.}
	\label{FIGHoLKor5}
\end{figure}

In figure~\ref{FIGHoLKor5}, we compare the self-diffusion coefficients $D_{1}$ of a hard sphere fluid as functions of $\eta_{1}$ in these two different matrices at different $\eta_{0}$.
As we can see, the self-diffusion coefficient $D_{1}$ in an overlapping hard sphere matrix at the same $\eta_{0}$ and $\eta_{1}$ for all cases is higher than in a hard sphere matrix and this difference increases with an increasing value of parameter $\eta_{0}$.

\section{Conclusions}

In this paper, we extended the Enskog theory to hard sphere fluids in disordered porous media and used it for the description of the self-diffusion coefficient of hard spheres. At the beginning, we  modelled a hard sphere fluid in a disordered porous medium by an equilibrium two-component mixture in the limit $m_{0}\to\infty$ for the species corresponding to the matrix particles. In such an approach, the contact values of the fluid-matrix and fluid-fluid pair distribution functions are introduced as the input of the Enskog theory. For their calculation, we used the scaled particle theory previously extended by us in \cite{HolDong09,ChenDong10,HolShmot10,PatHol11,HolPat12,HolPatDong12,HolPatDong17}  for the description of thermodynamic properties of hard sphere fluids in disordered porous media. The expressions obtained for the contact values of the fluid-matrix and fluid-fluid pair distribution functions are described only by the geometric porosity $\phi_{0}$ and do not include the dependence on the probe particle porosity $\phi$ and porosity $\phi^{*}$ defined by the maximum value of the fluid packing fraction of a hard sphere fluid in a porous medium. All three types of porosity are important for the description of thermodynamic properties of a hard sphere fluid in a disordered medium. We showed that the  application of such contact values neglects the effects of trapping of fluid particles by the matrix and at least the probe particle porosity $\phi$ should be included in the Enskog theory for a correct description of the matrix effect. Since the contact values of the fluid-fluid and fluid-matrix pair distribution functions have proven to be in very good agreement with computer simulations data \cite{KalHol14}, we consider that the Enskog theory or any other theory based only on static correlations between particles cannot correctly describe the effect of a static disordered matrix. Such a conclusion was also reached by Yethiraj with coworkers \cite{ChangJad04} from the analysis of the computer simulations data.

In the present paper, we extend the Enskog theory by modifying the contact values of the fluid-matrix and fluid-fluid pair distribution functions with new properties which include the dependence not only on  the geometric porosity $\phi_{0}$ but also on the probe particle porosity $\phi$. In this procedure, we consider that in the limit $\phi\to\phi_{0}$ these fictitious contact values coincide with the corresponding real contact values obtained by us in the framework of the scaled particle theory. We should note that the correction of the fluid-matrix contact value and the correction of the fluid-fluid contact value play different roles. The correction of the fluid-matrix contact value is very important for the description of the self-diffusion coefficient at small fluid densities and the correction of the fluid-fluid contact value can be important for the correction of the density dependence of the self-diffusion coefficient. Due to this, in this paper as the first step of such correction we consider only the modification of the fluid-matrix contact value. In the infinite dilution limit $\eta_{1}\to0$, we have the expression (\ref{HolKor3.5}) for the fluid-matrix contact value. Subsequently, at a finite value of $\eta_{1}$, we have the expression (\ref{HolKor3.7}) which corresponds to the approximation SPT2b for the description of thermodynamic properties \cite{PatHol11,HolPat12}. Then, we use the approximation (\ref{HolKor3.8}) which leads to the expression~(\ref{HolKor3.9}) corresponding to the approximation SP2b1 in the description of thermodynamic properties \cite{HolPat12,HolPatDong12,HolPatDong17}. In order to describe the fluid-fluid contact value, we left the expression (\ref{HolKor2.13}) without any modification. We will discuss the possibility of modification of $g_{11}(\sigma_{11})$ in a separate paper.

We showed that such a semi-empirical improvement of the Enskog theory predicts the correct trends for the influence of a porous medium on the diffusion coefficient of a hard sphere fluid. We obtained a good agreement with computer simulations data. We  discussed the effects of fluid density, fluid to matrix sphere size ratio, matrix porosity and matrix morphology on the self-diffusion coefficient. We also plan to consider some other transport coefficients such as shear and bulk viscosities. In addition, we intend to generalize the results obtained to binary hard sphere mixtures in porous media.

\section*{Acknowledgement} 
M.H. acknowledges support from the European Union’s Horizon 2020 research and innovation
programme under the Marie Sklodowska-Curie grant agreement No~734276.
M.H. also acknowledges support from the Ministry of Education and
Science of Ukraine (grant No. M/116-2019). 

We thank Ivan Kravtsiv for careful reading of the manuscript and useful comments.

\ukrainianpart

\title{Дифузія плину твердих сфер у невпорядкованому пористому середовищі. Застосування теорії Енского}

\author{М. Головко,  М. Корвацька }
\address{
Інститут фізики конденсованих систем НАН України, вул. Свєнціцького, 1,
79011 Львів, Україна
}

\makeukrtitle

\begin{abstract}
Ми застосовуємо теорію Енского для опису коефіцієнта самодифузії плину твердих сфер в невпорядкованому пористому середовищі. Використовуючи раніше розроблену нами теорію масштабної частинки для опису термодинамічних властивостей плину твердих сфер, прості аналітичні вирази для контактних значень плин-плин та плин-матриця парних функцій розподілу отримуються та використовуються як вхідні теорії Енского. Вирази, отримані для контактних значень, описуються лише геометричною пористістю і не включають залежність від інших типів пористості, важливих для опису термодинамічних властивостей. Показано, що застосування таких контактних значень нехтує ефектами захоплення частинок плину матрицею і, принаймні, термодинамічна пористість повинна бути включена в теорію Енского для правильного опису впливу матриці. У цій роботі ми розширюємо теорію Енского, змінюючи контактні значення парних функцій розподілу плин-матриця та плин-плин новими властивостями, які включають залежність не тільки від геометричної пористості, але й від термодинамічної пористості. Показано, що таке напівемпіричне вдосконалення теорії Енского відповідає наближенню SPT2b1 для опису термодинамічних властивостей і передбачає правильні тенденції впливу пористих середовищ на коефіцієнт дифузії плину твердих сфер в невпорядкованому пористому середовищі. Проілюстровано добре узгодження з комп'ютерними моделюваннями. Обговорюється вплив щільності плину, співвідношення розмірів плину твердих сфер та матриці, пористості матриці та морфології матриці на коефіцієнт самодифузії плину твердих сфер.

\keywords  плин твердих сфер, невпорядковані пористі середовища, теорія Енского, коефіцієнт самодифузії, теорія масштабної частинки, термодинамічна пористість
\end{abstract}


\begin{thebibliography}{10}
	%1
	\bibitem{GelbGub99} Gelb L.D., Gubbins K.E., Radhakrishnan R., Sliwinska-Bartkowiak M., Rep. Prog. Phys., 1999, {\bf62}, 1573, \\
	\bibdoi{10.1088/0034-4885/62/12/201}.
	%2
	\bibitem{MadGlandt88} Madden W.G., Glandt E.D., J. Stat. Phys.,  1988, {\bf51}, 537,
	\bibdoi{10.1007/BF01028471}.
	%3i
	\bibitem{ChangJad04}Chang R., Jagannathan K., Yethiraj A., Phys. Rev. E, 2004, {\bf69}, 051101,
	\bibdoi{10.1103/PhysRevE.69.051101}.
	%4i
	\bibitem{Krakov} Krakoviak V., Phys. Rev. E, 2007, {\bf 75}, 031503,
	\bibdoi{10.1103/PhysRevE.75.031503}.
	%5i
	\bibitem{KurCos11} Kurzidim J., Coslovich D., Kahl G., J. Phys.: Condens. Matter, 2011, {\bf 23}, 234122,\\
	\bibdoi{10.1088/0953-8984/23/23/234122}.
	%6
	\bibitem{GivenStall92}	Given J.A., Stell G., J. Chem. Phys., 1992, {\bf97}, 4573, \doi{10.1063/1.463883}.
	\bibitem{Ros99} Rosinberg M.-L., In: New Approaches to the Problems in Liquid State Theory,  Caccamo C., Hansen J.-P., Stell~G.~(Eds.),
	Kluwer, Dordrecht, 1999, 245--278,  \doi{10.1007/978-94-011-4564-0_13}.
	%8
	\bibitem{Pizio00} Pizio O., In: Computational Methods in Surface and Colloidal Science Series, Vol.~{89}, Borowko M. (Ed.) Kluwer,  Marcell Deker, New York, 2000, 293--346.
	%9
	\bibitem{Trokh96} Trokhymchuk A.D., Pizio O., Holovko M.F., Sokolowski S., J. Phys. Chem., 1996, {\bf100}, 17004,\\
		\bibdoi{10.1021/jp961443l}.
	%10
	\bibitem{Trokh97} Trokhymchuk A.D., Pizio O., Holovko M.F., Sokolowski S., J. Chem. Phys., 1997, {\bf106}, 200,\\
	\bibdoi{10.1063/1.473042}.
	%11
	\bibitem{HolDong09} Holovko M., Dong W., J. Phys. Chem. B, 2009, {\bf113}, 6360,
		\bibdoi{10.1021/jp809706n}.
	%12
	\bibitem{ReissFrisLeb59} Reiss H., Frisch H.L., Lebowitz J.L., J. Chem. Phys., 1959, {\bf 31}, 369,
	\bibdoi{10.1063/1.1730361}.
	%13
	\bibitem{ReissFrisHel60} Reiss H., Frisch H.L., Helfand E., Lebowitz J.L., J. Chem. Phys., 1960, {\bf 32}, 119,
	\bibdoi{10.1063/1.1700883}.
	%14
	\bibitem{ChenDong10} Chen W., Dong W., Holovko M., Chen X.S.,  J. Phys. Chem. B, 2010,  {\bf 114}, 1225,
		\bibdoi{10.1021/jp9106603}.
	%15
	\bibitem{HolShmot10} Holovko M.F., Shmotolokha V.I., Dong W., Condens. Matter Phys., 2010, {\bf 13}, 23607,\\
	\bibdoi{10.5488/CMP.13.23607}.
	%16
	\bibitem{PatHol11} Patsahan T., Holovko M., Dong W., J. Chem. Phys., 2011, {\bf 134}, 074503,
	\bibdoi{10.1063/1.3532546}.
	%17
	\bibitem{HolPat12} Holovko M., Patsahan T., Dong W., Pure Appl. Chem., 2012, {\bf 85}, 115,
		\bibdoi{10.1351/PAC-CON-12-05-06}.
	%18
	\bibitem{HolPatDong12} Holovko M., Patsahan T., Dong W., Condens. Matter Phys., 2012, {\bf 15}, 23607,
	\bibdoi{10.5488/CMP.15.23607}.
	%19
	\bibitem{HolPatDong17} Holovko M., Patsahan T., Dong W., Condens. Matter Phys., 2017, {\bf 20}, 33602,
	\bibdoi{10.5488/CMP.20.33602}.
	%20
	\bibitem{HolShmot14} Holovko M.F., Shmotolokha V.I., Patsahan T.,  J. Mol. Liq., 2014, {\bf 189}, 30,
		\bibdoi{10.1016/j.molliq.2013.05.030}.
	%21
	\bibitem{HolShmot18} Holovko M.F.,  Shmotolokha V.I., Condens. Matter Phys., 2018, {\bf 21}, 13602,
		\bibdoi{10.5488/CMP.21.13602}.
	%22
	\bibitem{ChenZhao16} Chen W., Zhao S.L., Holovko M., Chen X.S., Dong W., J. Phys. Chem. B,  2016, {\bf 120}, 5491,\\
	\bibdoi{10.1021/acs.jpcb.6b02957}.
	%23
	\bibitem{HvozdPat18} Hvozd M., Patsahan T., Holovko M., J. Phys. Chem. B,  2018, {\bf 122}, 5534,
	\bibdoi{10.1021/acs.jpcb.7b11834}.
	%24
	\bibitem{HolPat15} Holovko M., Patsahan T., Shmotolokha V., Condens. Matter Phys., 2015, {\bf 18}, 13607,\\
	\bibdoi{10.5488/CMP.18.13607}.
	%25
	\bibitem{HolShmot20} Holovko M., Shmotolokha V., Condens. Matter Phys., 2020, {\bf 23}, 13601, \doi{10.5488/cmp.23.13601}.
	%
	\bibitem{KalHol14} Kalyuzhnyi Yu.V., Holovko M., Patsahan T., Cummings P.T., J. Phys. Chem. Lett., 2014, {\bf 5}, 4260,\\
	\bibdoi{10.1021/jz502135f}.
	%27
	\bibitem{HolPatPat16} Holovko M.F., Patsahan O., Patsahan T., J. Phys.: Condens. Matter, 2016, {\bf 28}, 414003,\\
	\bibdoi{10.1088/0953-8984/28/41/414003}.
	%28
	\bibitem{HolPatPat17} Holovko M., Patsahan T., Patsahan O., J. Mol. Liq., 2017, {\bf 228}, 215,
	\bibdoi{10.1016/j.molliq.2016.10.045}.
	%29
	\bibitem{HolovPatPat17} Holovko M., Patsahan T., Patsahan O., J. Mol. Liq., 2017, {\bf 235}, 53,
	\bibdoi{10.1016/j.molliq.2016.11.030}.
	%30
	\bibitem{ResibLee77} Resibois P., de Leener M., Classical Kinetic Theory of Fluids, Wiley, New York, 1977.
	\bibitem{Boon80} Boon J.P., Yip S., Molecular Hydrodynamics, Dover, New York, 1980.
	\bibitem{McQuarrie76} McQuarrie D.A., Statistical Mechanics, Harper and Row, New York, 1976. 
	\bibitem{Chapman70} Enskog D.,  Kungl. Svenska Vet.-Ak. Handl., 1922, {\bf 63}, No.~1, 1--44 (in Swedish).
	\bibitem{YukhHol80} Yukhnovskii I.R., Holovko M.F., Statistical Theory of Classical Equilibrium Systems, Naukova Dumka, Kyiv, 1980, (in Russian).
	\bibitem{Minton} Minton A.P., J. Biol. Chem.,  2001, {\bf 276}, 10577,
	\bibdoi{10.1074/jbc.R100005200}.
\end{thebibliography}
\end{document}